\documentstyle[12pt,psfig]{article}
\begin{document}
\title{Instantaneous Inertial Frame but Retarded Electromagnetism 
in Rotating Relativistic Collapse}
\author{J. Katz$^{4,1}$, D. Lynden-Bell$^{1,2,3,4}$ and J. Bi\v{c}\'{a}k$^{5,1,4}$}
\maketitle
\centerline{Institute of Astronomy$^{1}$ and Clare College$^{2}$, Cambridge,}
\centerline{PPARC ``senior fellow'' on leave at School of Maths and Physics,}
\centerline{ The
Queens University$^{3}$, Belfast BT7 1NN}
\centerline{Racah Institute of Physics$^{4}$, the Hebrew University of Jerusalem}
\centerline{Dept of Theoretical Physics, Charles University$^{5}$, Prague}

\begin{abstract}

Slowly rotating collapsing spherical shells have flat spaces inside
and the inertial frames there rotate at $\omega_s(t)$ relative to
infinity.  As first shown by Lindblom \& Brill the inertial axes
within the shell {\it rotate rigidly without time delays} from one
point to another.  Although the rotation rate of the inertial axes is
changing the axes are inertial so, relative to them, there is neither
an ${\dot{\mbox {\boldmath $\omega$}}}_s \times {\bf r}$ (Euler)
fictitious force nor any other.  However, Euler and other fictitious
forces arise in the frame which is at rest with respect to infinity.
An observer at the center who looks in one direction $(\phi = 0, \;
\theta = \frac{\pi}{2}$ say) fixed to infinity will see that the sky
appears to rotate and can compare its apparent rotation with those of
the local inertial frame and of the shell itself.

By contrast in the electromagnetic analogue there is a time delay in the
propagation of the magnetic field inside a rotating collapsing charged
shell in flat space.

We demonstrate this time delay by devising a null experiment in which the
Larmor precession of a charged oscillator would be exactly cancelled by
the rotation of the inertial frame but for the delay.

In the \lq\lq combined" problem of a collapsing charged shell we show
that, due to the coupling of electromagnetic and gravitational
perturbations, the instantaneous rotation of inertial frames inside the
shell can be caused by pure electric currents in a non-rotating shell.
\end{abstract}

\section{Introduction}

Gravity is a tensor field and electromagnetism a vector field but
there are striking similarities between them.  At the classical level
there is the clear analogy between their inverse square force laws.
At the post-Newtonian level moving charges generate magnetic fields
while moving mass currents generate gravomagnetic\footnote{We have
adopted this more sonorous shorter form of Ciufolini \& Wheeler's
\cite{1} ``gravitomagnetic''.  For definitions of ${\bf B}_g$ in
strongly relativistic stationary space-times, see Lynden-Bell \&
Nouri-Zonoz \cite{15}.} fields.  These fields $B_g$ are the spatial
components of the curl of ${\mbox {\boldmath $\xi$}} / |{\mbox {\boldmath$\xi$}} |^2$,
${\mbox {\boldmath $\xi$}}$ being the time-like Killing vector.  When
the space-time is stationary, they give in different language a
description of the effects that are often called \lq\lq the dragging
of inertial frames" by relativists, but the advantages of the
gravomagnetic field concept have recently been stressed by Rindler
\cite{2}.

The main purpose of this paper is to demonstrate an interesting
difference in behavior between the gravomagnetic field which gives the
rotation of the inertial frame within a slowly rotating massive
collapsing spherical shell and the magnetic field within a rotating
collapsing shell of charge.  Whereas the gravomagnetic field has {\it
no time delay} and is uniform within the shell in the gravity problem,
nevertheless the magnetic field travels inwards from the shell, {\it
has a time delay} and is therefore non-uniform in the electromagnetic
problem.  The origin of this difference can be traced to the fact that
gravity as a tensor field does not carry any dipolar waves, so a
collapsing slowly rotating shell generates none of them.  Thus the
dipole contributions in gravity are governed by an \lq instantaneous'
equation not a wave equation.  The uniform ${\bf B}$ inside a rotating
shell is dipolar, and thus reacts instantaneously in a frame in which
its center is at rest.  In electricity the monopole contributions are
governed by the instantaneous Gauss's theorem and remarkably that
instantaneous relationship between $\int{\bf E}\cdot d {\bf S}$ and
the included charge holds in all Lorentz frames!  However in
electromagnetism there are dipolar waves and a collapsing rotating
spherical shell generates them.  The increase in ${\bf B}$ as the
shell collapses and rotates faster travels inwards from the shell.
This difference in the time delay behavior has therefore a good
physical origin, nevertheless, all physicists shy away from
instantaneous effects as acausal.  How can it be that the internal
inertial frames rotate at once without waiting for a causal signal to
travel inwards from the shell?

Whereas local inertial frames can be determined locally their rotations
and accelerations relative to inertial frames at infinity depend on a
convention of how such frames are extended inwards.  There is nothing
wrong with conventions that travel with the \lq speed of thought'.  One
may consider the Andromeda Nebula at one moment and Jerusalem a second
later, so thoughts can travel far faster than light.  Thus, when we find
that the inertial frame within a collapsing rotating spherical shell
rotates rigidly with no time delay between the center and the shell's
inner edge, we should not claim some spooky faster than light influence,
rather we should look at the way we define the rotation of a reference
frame.  It is we who decide on the \lq gauge' in which the rotation is
measured - it is our thoughts that travel faster than light!

Our metric is stationary outside the shell with coordinates $t, \phi$ such that observers at rest at infinity have $t$ as their proper time and remain with $\phi$ constant.  We extend this $\phi$ continuously into the shell's flat interior and with the appropriate time the inertial frame there rotates rigidly through the surfaces of constant $\phi$.  The angular velocity is directly related to the observable effects analysed by Lindblom \& Brill \cite{3}.

In the time-slicing we use, the rotation of inertial frames is determined
by the constraint equations and it is thus related to instantaneous
values of the matter variables.

\section{Principal results}

In this section we give the essence of what we have done.  Proofs are
given in the sections that follow.  Familiar notations are not defined in
this section; units are those for which $G=c=1$. 

\subsection{Motion of inertial frames and fixed stars as seen from the 
center of a collapsing shell (Section 3)}

A collapsing spherical shell of dust in slow rotation produces a
slightly perturbed Schwarzschild spacetime which in spherical
coordinates $x^{\mu} \equiv (x^o = t,\break \; x^1 = r, \; x^2 = \theta, \;
x^3 = \varphi)$ has a metric of the following form outside the shell
$(r \geq r_s)$ -- see Lindblom and Brill \cite{3}:
$$ds^2 = \left( 1 - \frac{2M}{r}\right) dt^2 -
\frac{1}{1-\frac{2M}{r}} dr^2 - r^2 d\theta^2 - r^2{\rm sin}^{2} \theta
(d\varphi - \omega dt)^2 , \ \ \ \ \  
r \geq r_s \ . \eqno (2.1)$$

The perturbation $\omega (r)$ is due to the slow rotation of
the shell and
is proportional to the fixed total angular momentum $J$:
$$\omega = \frac{2J}{r^3} \ . \eqno (2.2)$$
On the shell itself, $r = r_s(t),$ and $\omega$ is a
function of time
$$\omega [r_s(t)] = \frac{2J}{r_s^3} \equiv \omega_s \ . \eqno (2.3)$$
Note that $\omega_s$ is not the rotation of the shell, this will
be denoted by a capital omega.  The flat metric inside $(r \leq r_s)$,
written in spherical coordinates $\overline{x}^{\mu} \equiv (\overline{x}^{o} =
\overline{t},\; \overline{x}^{1} = r, \;\overline{x}^2 = \theta,\;
\overline{x}^3 = \overline{\varphi})$ is
$$d\overline{s}^2 = d\overline{t}^2 - dr^2 - r^2 d\theta^2 - r^2\; {\rm
sin}^2 \theta d\overline{\varphi}^2 \, , \qquad r \leq r_s \ . \eqno (2.4)$$
The matching of the metrics (2.1) and (2.4) on the shell tells
us how
$\overline{t}$ is related to $t$ [the ratio of (3.5) and (3.6)] and
$\overline{\varphi}$ to $\varphi$:
$$ \overline{\varphi} = \varphi - \int^t \; \omega_s dt \ . \eqno
(2.5)$$ We see therefore from (2.4) and (2.5) that the local inertial
frames inside ($\overline{\varphi} =$ const.) all rotate rigidly with
the same angular velocity with respect to observers at rest relative
to infinity ($\varphi =$ const.):
$$\frac{d\overline{\varphi}}{dt} = 0 = > \frac{d\varphi}{dt} = \omega_s \ . \eqno (2.6)$$
Observers at constant $r, \theta, \varphi$ we refer to as static
observers.
They are not inertial.  The rotation of inertial frames inside rotating
non-collapsing shells was recognized long ago by Thirring \cite{4} and has been
studied by Brill and Cohen \cite{5} and others (see Barbour and Pfister \cite{6}).
For expanding or collapsing shells $\omega$ depends on $t$, but the
internal inertial frames still rotate rigidly -- a remarkable result first
found by Lindblom and Brill \cite{3}.  This was generalized to a cosmological
treatment of Mach's principle by us \cite{7}.   An inertial observer anywhere
inside would preferably use his own proper time $\overline{t}$ to
calculate his rate of rotation with respect to infinity.  The rate of
rotation, 
$$\frac{d\varphi}{d\overline{t}} = \overline{\omega}_s \ ,  \eqno (2.7)$$
is the ``dragging'' coefficient on the shell.  

The function $\omega(r)$ defined in (2.2) is simply related to the rate of
rotation $\overline{\Omega}_{\tau}$ of the shell itself measured in units
of its proper comoving time $\tau$:
$$\overline{\Omega}_{\tau} = \frac{d\overline{\varphi}_{\rm
shell}}{d\tau} \ . \eqno (2.8)$$
If $m_s$ represents the constant proper mass of a shell of dust,
$$ \overline{\Omega}_{\tau} = \frac{1}{4m_s} \;
\left . \frac{\partial\omega} {\partial (1/r)} \right | _{r=r_{s}} \; = \;
\frac{3r_s}{4m_s} \; \omega_s = \frac{3J}{2m_s r_s^2} \ . \eqno (2.9) $$
The first equality is independent of the detailed structure of
$\omega$.  

Lindblom and Brill have analyzed various consequences of the rotation of
inertial frames. In Section 3 we first show that covariantly defined
acceleration
of static observers inside is simply given by $\mid
\frac{d\overline{\omega}_s}{d\overline{t}} \mid$, and the magnitude of the
(covariantly defined) vorticity of their worldlines is $\mid
\overline{\omega}_s\mid$.   These results can also easily be obtained by
means of the \lq\lq gravitational vector potential" which arises in the
non-inertial frame connected with those observers.  
$\left | \frac{d\overline{\mbox {\boldmath
$\omega$}}_s}{d\overline{t}} \times {\bf r} \right | $ is the usual
Euler fictitious acceleration.

Then we calculate the motion of fixed stars as seen by an inertial
observer at the center of a collapsing transparent shell in slow rotation.

Photons emitted radially inwards in the equatorial plane from a fixed
star at infinity appear to him to rotate around the z-axis with an angular
velocity
$$\overline{\omega}_{\rm star} \equiv \frac{d\overline{\varphi}_{\rm
star}}{d\overline{t}} = - \left( \overline{\omega}_s\;
\frac{1+V}{1+\overline{V}} \right)_{\rm ret} \ . \eqno (2.10)$$
$V$ is the proper radial velocity of the shell as measured by
static
observers outside,
$$V = {1 \over 1 - 2M/r_s} \;\; \frac{dr_s}{dt} \ , \eqno (2.11)$$
while 
$$\overline{V} = \frac{dr_s}{d\overline{t}} \eqno (2.12)$$
is the proper velocity measured in time $\overline{t}$ from the
inside.
The right hand side in (2.12) must be evaluated at a retarded time
$\overline{t}_{\rm ret}$, the moment the photon crossed the shell:
$$\overline{t}_{\rm ret} = \overline{t} - r_s(\overline{t}_{\rm ret})
\ . \eqno (2.13)$$ 
Section 3 is mainly devoted to the derivation of
(2.10).  If the shell's radius is fixed, $V = \overline{V} =0$, and the
shell is in uniform rotation, $\overline{\omega}_s =
\overline{\omega}_{s0}$, then
$$\overline{\omega}_{\rm star} = - \overline{\omega}_{s0} \ . \eqno
(2.14)$$ Fixed stars appear to rotate (backwards) with an angular
velocity $-\overline{\omega}_{s0}$ just as static inside observers do.
With shells of changing radius $\mid V \mid$ is always greater than
$\mid \overline{V} \mid$.  Thus if the shell collapses ($V<0$), fixed
stars appear to rotate more slowly because $\overline{\omega}_{\rm
star} > - \overline{\omega}_s$.  In the limit, when the shell moves
with a radial velocity close to that of light, $\overline{V}, V
\rightarrow -1$,
$$\overline{\omega}_{\rm star} \rightarrow - \left( 1 - \frac{2M}{r_s}
\right) \;
\overline{\omega}_{s{\rm ret}} \qquad \qquad (V, \overline{V} \rightarrow
-1) \ . \eqno (2.15)$$
With expanding shells $(V>0)$, the stars appear to rotate faster, 
$\overline{\omega}_{\rm star} < - \overline{\omega}_{s{\rm ret}}$, and if
$V, \overline{V} \rightarrow + 1$
$$\overline{\omega}_{\rm star} \rightarrow - \overline{\omega}_{s{\rm
ret}} \qquad \qquad (V, \overline{V} \rightarrow 1)\ . \eqno (2.16)$$

In summary, for an inertial observer inside the collapsing shell (who
rotates as seen from infinity): 

\begin{itemize}
\item[$(\overline{1})$]
The fixed stars are rotating (backwards) with angular velocity
$$ \frac{d\overline{\varphi}_{\rm star}}{d\overline{t}} \; = \;
\overline{\omega}_{\rm star} = - \frac{1+\frac{m_s}{M} {\dot r}_s -
\frac{m_s^2}{2r_{s}M}}{1+\frac{m_s}{M}\dot{r}_s + \frac{m_s^2}{2r_{s}M}}
\;\; \left . \frac{\omega_s}{1 - \frac{2M}{r_s}} \right| _{\overline{t}_{\rm ret}} \ . \eqno (2.17)$$
\item[$(\overline{2})$]
The shell is rotating (forwards) with angular velocity 
$$\frac{d\overline{\varphi}_{\rm shell}}{d\overline{t}} = \; 
\overline{\Omega} \; = \frac{3r_s}{4M} \; \frac{\omega_s}{1 + \frac{m_s^2}{2r_sM}} \ . \eqno (2.18)$$
\item[$(\overline{3})$]
The spacelike geodesics (e.g., $t = \ {\rm const.,} \  \varphi = 0, \ \theta = \frac{\pi}{2}$)
connected to fixed points at infinity are rotating (backwards) at a rate
$$\frac{d\overline{\varphi}}{d\overline{t}} = - \overline{\omega}_s \ . \eqno (2.19)$$
\end{itemize}

If instead of describing the world picture as it appears to those
observers we use world maps fixed at infinity and the universal time
$t$, then an observer inside who looks in a fixed direction in the map
sees:

\begin{itemize}
\item[(1)]
The fixed stars rotating (forwards) with angular velocity
$$\displaylines{\frac{d\varphi_{\rm star}}{dt} = \omega_{\rm star} =
(\overline{\omega}_{\rm star} + \overline{\omega}_s)
\frac{d\overline{t}_s}{dt_s}  \cr \hfill  = 
(\overline{\omega}_{\rm star} + \overline{\omega}_s)
\left(1 - \frac{2M}{r_s}\right) \;  \frac{1+\frac{m_s^2}{2r_sM}}{1-\frac{m_s^2}{2r_sM}} \ . \hfill {(2.20)}\cr}
$$
\item[(2)]
The shell rotating (forwards) with angular velocity 
$$\frac{d\varphi_{\rm shell}}{dt} \; = \Omega \; = \; \left(1 +
\frac{3r_s}{4M} \; \frac{1-\frac{2M}{r_s}}{1-\frac{m_s^2}{2r_sM}} \right)
\omega_s \ . \eqno (2.21)$$
\item[(3)]
The local inertial frame rotating (forwards) with angular velocity
$$\frac{d\varphi}{dt} = \omega_s = \frac{2J}{r_s^3} \ . \eqno (2.22)$$

\end{itemize}
Formulas (2.17) to (2.22) have been obtained by simple algebraic
combinations of formulas given in Section 3.  

Finally, if the shell is in steady rotation and not collapsing ($r_s =
r_{s0}$), a star at an angle $\varphi_{\infty}$ will be seen from the
center in the direction $\varphi_{\rm star 0}$ which, following
(3.27), is given by
$$\varphi_{{\rm star}_0} = \varphi_{\infty} - \frac{J}{2M^2} \left[
\frac{2M}{r_{s0}} \left(1 - \frac{2M}{r_{s0}}\right) + {\rm ln}
\left(1-\frac{2M}{r_{s0}}\right) \right] \ ; \eqno (2.23)$$
if the shell has a big radius compared to $M, \; 2M/r_s \ll 1$,
$$\varphi_{{\rm star}_0} \simeq \varphi_{\infty} + {\textstyle {3 \over 2}} \; \omega_{s0}
r_{s0} \ . \eqno (2.24)$$

\subsection{The electromagnetic problem in a flat spacetime (Section 4)}

The magnetic field ${\bf B}_0 = B_0{\hat{\bf z}}$ of a shell of
constant radius $r_{s0}$ with a uniformly distributed charge $Q$ that
rotates with a small constant angular velocity $\Omega_0$ is uniform
inside:
$$B_0 = \frac{2}{3} \frac{Q}{r_{s0}} {\overline {\Omega}}_0  \qquad \qquad {\rm
(stationary)} \ . \eqno (2.25) $$

If the shell starts to collapse at $\overline{t} = 0$, both the magnetic
and electric fields within the shell are determined by the vector
potential ${\bf A}$ which is of the form 

$${\bf A} = \frac{Q}{4r^3} \; {\mbox {\boldmath $\psi$}} \times {\bf r} \ , \eqno (2.26)$$
with
$${\mbox {\boldmath $\psi$}} ({\overline t}, r) = -
\int_{{\overline t}^-}^{\overline {t}^+} \,  {\overline{{\mbox 
{\boldmath $\Omega$}}}  (t') \over r_s (t') \sqrt{1-{\overline V} (t')^{2}}} \, 
\left[({\overline t} - t^{\prime})^2 -r^2 - r_s^{\prime 2} \right]
dt^{\prime}\ . \eqno {(2.27)}$$ 
Here ${\overline t}^- ({\overline t},
r)$ and ${\overline t}^+ ({\overline t}, r)$ are the retarded and
advanced times defined implicitly by
$$\overline{t}^- = \overline{t} - r - r_s (\overline{t}^-) \qquad , \qquad 
\overline{t}^+ = \overline{t} + r - r_s (\overline{t}^+) \ . \eqno (2.28)$$
At the center
$$\overline{t}^- = \overline{t}^+ \; = \; \overline{t}_{\rm ret} \; = \;
\overline{t} - r_s (\overline{t}_{\rm ret}) \eqno (2.29)$$
like in (2.13).

The potential outside $(r > r_s)$ is given by the same formula (2.26), (2.27)
except that there ${\overline t}^+$ is to be replaced by
$$\overline{t}^* \; = \; \overline{t} - r  + r_s (t^*) \ . \eqno (2.30) $$

Electromagnetic fields ${\bf E}, {\bf B}$ inside a collapsing shell
are not uniform. Near the center $C$, the lowest orders are respectively
given by 
$${\bf B} = {\bf B}_{C} (\overline{t}) + r^2 {\bf B}_2 + \ldots \ , \eqno (2.31)$$
$${\bf E} = - \frac{1}{2} \; \frac{d{\bf B}_C}{d\overline{t}} \times
{\bf r} + r^2 {\bf E}_2 \times {\bf r} + \ldots \ , \eqno (2.32)$$
in which ${\bf B}_C = B_C {\hat{\bf z}}$ where
$$B_C = \frac{2}{3} \; Q \left[ \frac{1}{1+\overline{V}} \;
\frac{d}{d\overline{t}} \; \left(\frac{1}{1+\overline{V}} \;
\frac{\overline{\Omega}}{\sqrt{1-\overline{V}^2}} \right) \; + \;
\frac{1}{(1+{\overline V})^2} \;
\frac{\overline{\Omega}/r_s}{\sqrt{1-\overline{V}^2}}
\right]_{\overline{t} \rightarrow \overline{t}_{\rm ret}} \ . \eqno (2.33)$$
This is a function of the retarded time defined by (2.13).

Section 4 and the Appendix are mainly devoted to deriving Eqs. (2.27) and (2.33), which are
used to analyze the motion of a particle.

\subsection{Motion of a particle inside a charged collapsing rotating shell}

We now consider a charged massive shell in slow rotation and fast
collapse which brings us back to general relativity.  The metric outside
becomes a perturbed Reissner Nordstr\"om metric.  The line element is of
the form (2.1) with $1-\frac{2M}{r}$ replaced by $1 - \frac{2M}{r} +
\frac{Q}{r^2}$.  The function $\omega$ is no longer given by (2.2) and
depends on ${\overline t}$ as well as $r$.  

In the charged case, not only is the rotation of the material shell the
source of $\omega$ but any dipole odd-parity electromagnetic field (e.g., a current loop in the equatorial plane) will
contribute also. The interacting electromagnetic and gravitational
perturbations of Reissner-Nordstr\"om spacetime were analyzed by
Bi\v{c}\'{a}k \cite{8} in detail.  In the case of odd-parity
dipole perturbations, the function $\omega(r, {\overline t})$ is given (cf. the equation
below Eq. (59) and Eq. (60) therein) by the relation
$$ \omega (r, {\overline t}) = \frac{2J}{r^3}\left( 1 - {Q^2 \over 2Mr}\right) + \sqrt{{3 \over 4 \pi}}  \int ^\infty _r  \; \frac{2Q}{r'^4} \;
{\cal P}_f ({\overline t}, r') dr' \ , \eqno (2.34)$$
${\cal P} _f ({\overline t},r)$ describes a dipole electromagnetic field of
odd parity.  Owing to the non-vanishing background electric field outside
the shell, the electromagnetic and gravitational perturbations are
coupled.  The small dipole electromagnetic perturbation described by
${\cal P}_f$ thus becomes, according to (2.34), the source of
gravitational dragging.  Even if the shell is neither rotating nor collapsing, an electric
current on the shell will be the source of a dragging even though the
matter of the shell is at rest,\footnote{See Bi\v{c}\'{a}k
\& Dvo\v{r}\'{a}k \cite{9} for the detailed discussion of such
interacting stationary electromagnetic and gravitational perturbations.} with respect to static observers at infinity.

If the slowly rotating, charged shell is collapsing, the
time-dependent electromagnetic field outside the shell will be the
source of a time-dependent $\omega(r,{\overline t})$ according to
(2.34).  The electromagnetic waves outside the shell will also
slightly backscatter off the curvature of spacetime when the shell
becomes relativistic and can even penetrate inside the shell's flat
interior.  Since these wave tails are much weaker than the primary
waves from the shell, we shall neglect them.  In our $(t, r, \theta,
\phi)$ coordinates, the electromagnetic field inside the shell is then
the same as that discussed above for a collapsing charged shell in
Minkowski space in $(t, r, \theta, \phi)$ coordinates.  (Actually,
provided that we measure the current and the field in the same axes
they bear the same relationship to each other in $({\overline t}, r,
\theta, {\overline \phi})$ coordinates too, provided we neglect terms
of order omega squared.)  By joining the outside perturbed
Reissner-Nordstr\"om metric across the shell with the flat inside, we
obtain the factor $\omega (r_s(\overline t), \overline t) \equiv
\omega_s(\overline t)$ inside the shell, as before in the uncharged
case.

In inertial coordinates $(\overline{t}, {\bf r})$ the
equations of motion of a particle of mass $m$ and charge $q$ near the center
and subject to the electromagnetic field (2.31) (2.32) are, in the presence of a $-k {\bf r}$ restoring force 
$$m\frac{d^2 {\bf r}}{d\overline{t}^2} \; = \; q\left( -
\frac{1}{2} \;
\frac{d {\bf B}}{d\overline{t}} \times {\bf r} + {d {\bf r} \over d {\overline t}} \times
{\bf B} \right) - k {\bf r} \ . \eqno (2.35) $$
Let us introduce the Larmor precession vector
$$ {\overline{\mbox {\boldmath $\omega$}}}_L \; = \; \frac{q {\bf B}}{2m}\ . \eqno (2.36)  $$
Then
$$
\frac{d^2 {\bf r}}{d\overline{t}^2} \; = \; - \frac{d{\overline{\mbox
{\boldmath $\omega$}}}_L}{d\overline{t}} \; \times \; {\bf r} - 2
{\overline{\mbox {\boldmath $\omega$}}}_L \times {d {\bf r} \over d
{\overline t}} \ . \eqno (2.37)$$ In our inertial frame, static
coordinates rotate at angular velocity $-\overline{\omega}_s$ [see
Eq. (2.19)].  With respect to fixed static non-inertial coordinates,
suffix $f$, $\frac{d{\bf r}}{d\overline{t}}$ goes to ${\bf v} -
{\overline{\mbox {\boldmath $\omega$}}}_s \times {\bf r}$ and the
acceleration becomes
$$
\left. \frac{d^2{\bf r}}{d\overline{t}^2} \right | _f - \left . \frac{d{\overline{\mbox
{\boldmath $\omega$}}}_s}{d\overline{t}} \right | _f \times  {\bf r} - 2
{\overline{\mbox {\boldmath $\omega$}}}_s \times {\bf v} = -\frac{d
{\overline{\mbox {\boldmath $\omega$}}}_L}{d\overline{t}} \; \times \;
{\bf r} - 2 {\overline{\mbox {\boldmath $\omega$}}}_L\; \times \; {\bf
v}\ , \eqno (2.38)$$ and centrifugal accelerations ${\overline{\mbox
{\boldmath $\omega$}}}_s \times ({\overline{\mbox {\boldmath
$\omega$}}}_s \times {\bf r})$, etc., which we neglect since they are
proportional to $\overline{\omega}^2$.  Thus, in coordinates fixed
at infinity, the equations of motion of the particle turn out to be, dropping the suffices $f$ and neglecting $\omega^2$ terms,
$$\frac{d^2{\bf r}}{d\overline{t}^2} = - {\bf r} \times
\frac{d}{d\overline{t}} \left( {\overline{\mbox {\boldmath
$\omega$}}}_s - {\overline{\mbox {\boldmath $\omega$}}}_L \right) - 2
{\bf v} \times \left( {\overline{\mbox {\boldmath $\omega$}}}_s -
{\overline{\mbox {\boldmath $\omega$}}}_L \right) - k {\bf r} \ . \eqno (2.39)$$
We may choose a linear oscillator with the particle mass and charge in such a way that before
the shell started to collapse,
$$\overline{\omega}_{L0} \; = \; \overline{\omega}_{s0} \ , \eqno (2.40)$$
resulting in an oscillator which oscillates without precession.  So long as ${\overline {\omega}}_s$ and $\overline {\omega}_L$ remain equal (2.39) ensures no precession.  However in (2.39) ${\overline {\mbox {\boldmath $\omega$}}}_L$ depends on $\overline{t}_{{\rm ret}}$ while $\overline {\omega}_s$ is evaluated at the non-retarded $\overline t$.  Thus as the infall gathers speed, ${\overline t}_{{\rm ret}}$ can no longer be approximated by $\overline t$ so the oscillator starts to precess relative to infinity-fixed axes.  Actually in the strongly relativistic r\'egime the formulae for ${\overline {\mbox {\boldmath $\omega$}}}_L$ and ${\overline {\mbox {\boldmath $\omega$}}}_s$ in terms of the sphere's rotation rate differ, so this is only a good null experiment in the weak-dragging r\'egime.

\section{Inertial frames and apparent motion of fixed stars}

\subsection{Properties associated with the radial motion of the shell}

Consider a thin shell with the metric outside given by (2.1) and inside
by (2.4) with (2.5).  Let $r=r_s(\tau)$ be the radius of the shell given
as a function of its proper time $\tau$.  Let $t=t_s (\tau)$ and
$\overline{t} = \overline{t}_s(\tau)$ represent the $t$ and
$\overline{t}$ dependence on $\tau$ on the shell.  By replacing $(t,r)$ by
($t_s, r_s)$ in (2.1) or $(\overline{t}, r)$ by $(\overline{t}_s, r_s)$ in
(2.4) we must obtain the same expression because it is the intrinsic
metric of the shell, namely,
$$ds_s^2 = \gamma_{ab}\, d\theta^a \,d\theta^b = d\tau^2 - r_s^2(\tau)
\left[d\theta^2 + {\rm sin}^2\theta \, d\overline{\varphi}^2 \right] \ . \eqno (3.1)$$
Consequently (2.1) implies a relation between $t_s$ and $r_s$: 
$$\left(1 - \frac{2M}{r_s}\right) \dot{t}_s^2 - \frac{\dot{r}_s^2}{1-
\frac{2M}{r_s}} = 1 \ , \eqno (3.2)$$
and (2.4) implies a relation between $\overline{t}_s$  and $r_s$,
$$\dot{\overline{t}}_s^2 - \dot{r}_s^2 = 1 \ . \eqno (3.3)$$ Dots
denote derivatives with respect to $\tau$ only.  From (3.2) and (3.3)
there follows a relationship between $\overline{t}_s$ and $t_s$. But
this is better given by ratio of (3.5) and (3.6) given below.
Subsequently, there will appear other time derivatives.

The equation of motion for the radius of the shell is derived by
integrating Einstein's (0,0) equation across it as given in Israel \cite{10}:
$$1 + \dot{r}_s^2 = \left(\frac{M}{m_s} \; + \; \frac{m_s}{2r_s}
\right)^2 \ , \eqno (3.4)$$
in which $m_s$ represents the proper mass of the shell of
uniform surface
density $\sigma = m_s/4\pi r_s^2$.  If we use (3.4) to eliminate
$\dot{r}_s$ from (3.2) and (3.3) and define $\dot{t}_s$ and
$\dot{\overline{t}}_s$ to be positive, we obtain the following useful
expressions for $\dot{t}_s$ and $\dot{\overline{t}}_s$: 
$$\dot{t}_s = \left(\frac{M}{m_s} \; - \; \frac{m_s}{2r_s}
\right) / \left( 1 - \frac{2M}{r_s} \right) \eqno (3.5) $$
and
$$\dot{\overline{t}}_s = \frac{M}{m_s} \; + \; \frac{m_s}{2r_s} \
. \eqno (3.6)$$

\subsection{The angular velocity of the shell and the inertial frames}

A spherical shell of dust with rest mass $m_s$ and local velocities
$w^a$
must have a surface energy tensor $\tau_b^a$ $(a, b, \ldots = 0, 2, 3)$ of the following form
$$\tau^a_b = \frac{m_s}{4\pi r_s^2} \; w_b \, w^a \qquad , \qquad
\gamma_{ab} w^a \, w^b = 1 \ , \eqno (3.7)$$ 
where $\gamma_{ab}$ is defined in (3.1). If the shell is slowly
rotating with proper angular velocity ${\overline \Omega}_{\tau}$,
$$ \overline{\Omega}_{\tau} = \frac{ d\overline{\varphi}_{\rm
shell}}{d\tau} \ , \eqno (3.8)$$
then $w^0 \simeq 1, \; w^2 = 0, \; w^3 \simeq
\overline{\Omega}_{\tau}$, since $r^2\overline{\Omega}_{\tau}^2 \ll 1$ is
neglected. The number of non-zero components of $\tau^a_b$ is reduced to
two:
$$ \tau_0^0 = \frac{m_s}{4\pi r_s^2} \qquad , \qquad \tau_{3}^0 = -
\frac{1}{4\pi}\; m_s \overline{\Omega}_{\tau} {\rm sin}^2 \theta \
. \eqno (3.9)$$
As a consequence of Einstein's equations there exists a simple relation between the components of the
energy tensor $\tau_b^a$ and the external curvature tensor components
$K^a_b, \;  \overline{K}^a_b$ on both sides of the shell.  The present
notations are close to those of the paper of Goldwirth and Katz \cite{11} to
which here we refer for details.  In particular, to our order of
approximation, it may be calculated that
$$\tau_3^0 = -\frac{1}{8\pi} \, K_3^0  =  -\frac{1}{16\pi} \;
\left . \frac{d\omega}{d(1/r)} \right | _s \sin^2\theta \ . \eqno (3.10)$$
Comparing the two expressions (3.9) and (3.10) for $\tau_3^0$,
we thus find that 
$$\overline{\Omega}_{\tau} = \frac{1}{4m_s}\; \left . 
\frac{d\omega}{d(1/r)} \right | _s = {3 r_s \over 4m_s} \, \omega_s\ . \eqno (3.11)$$
It is worthwhile to note that the first equality is independent of
the detailed structure of $\omega$ and holds thus also for slowly rotating
{\it charged} collapsing shell although $m_s$ will then be changing during the collapse.  Equation (3.11) for
$\overline{\Omega}_{\tau}$ has been given in (2.9).  Inside and outside
observers may, however, be inclined to use their own local times.  Thus,
the angular velocity of the shell $\overline{\Omega}$ in $\overline{t}$
time is - using (3.6):
$$\overline{\Omega} = \frac{d\overline{\varphi}_{\rm
shell}}{d\overline{t}}
\; = \; \frac{\overline{\Omega}_{\tau}}{\dot{\overline{t}}_s} \; = \;
\frac{3r_s/4m_s}{\frac{M}{m_s} + \frac{m_s}{2r_s}} \; \omega_s \
. \eqno (3.12)$$
For an observer outside, the angular velocity $\Omega$ is
related to
$\overline{\Omega}_{\tau}$ in a more complicated way:
$$ \frac{\overline{\Omega}_{\tau}}{\dot{t}_s} \; = \;
\frac{d\overline{\varphi}_{\rm shell}}{dt} \; = \; \frac{d\varphi_{\rm
shell}}{dt} - \omega_s = \Omega-\omega_s \ . \eqno (3.13)$$
So,
$$\Omega = \left[ \frac{(3r_s/4m_s)(1-2M/r_s)}{\frac{M}{m_s} -
\frac{m_s}{2r_s}} + 1 \right] \omega_s \ , \eqno (3.14)$$
or in Lindblom and Brill's form but {\it not} in their
notation,
$$ \omega_s = \frac{4M}{3r_s} \; (\Omega-\omega_s) \;
\frac{1-m_s^2/2Mr_s}{1-2M/r_s} \ . \eqno (3.15)$$
This expression has been given by Lindblom and Brill in
isotropic coordinates, with our $\omega_s$ written $\Omega$ and our
$\Omega$ written $\omega$. 

The time-dependent rigid rotation of inertial frames inside the shell
can well be illustrated by considering static observers.  They
experience Euler acceleration (their Coriolis and centrifugal
accelerations being negligible under our assumption of slow rotation),
and the congruence of their world lines twists.  Neglecting the terms
proportional to $\overline{\omega}_s^2,$ their four-velocity
$\overline{U}^{\mu}$ has two non-zero components; with (2.5) and
$\varphi$ = const.,
$$\overline{U}^0 = 1, \qquad \qquad \overline{U}^3 =
\frac{d\overline{\varphi}}{d\overline{t}} = - \overline{\omega}_s \
. \eqno (3.16)$$
The four-acceleration is
$$\overline{a}^{\mu} = \overline{U}^{\nu} \; \overline{D}_{\nu} \;
\overline{U}^{\mu} \ , \eqno (3.17)$$ where $\overline{D}_{\nu}$ is a
covariant derivative.  $\overline{a}^{\mu}$ has only one non-vanishing
component equal to
$$\overline a^3 = \frac{d\overline{\omega}_s}{d\overline{t}} \ . \eqno (3.18) $$

The acceleration (3.17) is equal to the physical acceleration of the
particle with $\overline{U}^{\mu}$ measured in the inertial frame in which
the particle is momentarily at rest.  It is easy to see that in our
approximation this is equal to the physical acceleration with respect to
our inertial frame inside.

Clearly, the acceleration (3.17), being given by an explicitly
covariant form, can be calculated in any system of coordinates.  It is
instructive to find it in a more \lq physical' way by calculating the
force on a particle at rest in a non-inertial frame with metric (2.4)
rewritten in terms of $\varphi$,
$$d\overline{s}^2 = d\overline{t}^2 - dr^2 - r^2d\theta^2 - r^2{\rm
sin}^2\theta (d \varphi - {\overline \omega}_s d{\overline t})^2 \ ,  \eqno (3.19)$$
in which static observers are at rest.

Using the formalism of (non-covariant) gravitational potentials (see, e.g., 
M\o ller \cite{13} and Zel'manov \cite{14}), we find that the vector of the field of
gravitational-inertial forces acting on the particle has only a
non-vanishing azimuthal component
$$\overline{F}_3^{(g)} = - \frac{\partial\overline{A}_3^{(g)}}{\partial
\overline{t}} = r^2 {\rm sin}^2 \theta \;
\frac{d\overline{\omega}_s}{d\overline{t}} \ , \eqno (3.20) $$
where $\overline{A}_i^{(g)} = -
\overline{g}_{0i}/\overline{g}_{00}$ is the gravitational vector potential
in $(\overline{t}, r, \theta, \varphi)$ coordinates.  This is the standard
Euler acceleration, observed in a non-inertial frame which is rotating
with a time-dependent angular velocity.  As expected, we find
$$\overline{F}_3^{(g)} = - \overline{a}_3 = r^2 {\rm sin}^2 \theta\;
\overline{a}^3 \ , \eqno (3.21)$$
with $\overline{a}^3$ given by (3.18).

Both Coriolis and centrifugal accelerations of the static observers are
proportional to $\overline{\omega}_s^2$.  However, a particle moving with
a general velocity inside the shell experiences a Coriolis acceleration
as observed in the static frame (3.19).

The twist (vorticity) of the congruence of timelike lines with unit
tangent vectors $\overline{U}^{\mu}$ is covariantly described by the
vorticity 4-vector (Misner {\it et al}. \cite{12})
$$\overline{\omega}^{\mu} = \overline{\epsilon}^{\mu\nu\rho\sigma}
\overline{U}_\nu (\delta_\rho ^{\lambda} - \overline{U}_\rho
\overline{U}^{\lambda}) \overline{D}_{\lambda} \overline{U}_{\sigma} \
. \eqno (3.22)$$ With $\overline{U}^{\mu}$ given by (3.16) and the
metric (3.19) we obtain the non-vanishing components
$$\overline{\omega}^{\mu} = (0, \overline{\omega}^j) = (0,
-\overline{\omega}_{s}\cos  \theta, \; \frac{\overline{\omega}_s}{r}
\,{\rm
sin} \theta, 0) \eqno (3.23)$$
or, in $\overline{{\bf r}}$ coordinates,
$\overline{\omega}^{\mu}
= (0, {\overline{\mbox {\boldmath $\omega$}}})$ with 
$${\overline{\mbox {\boldmath $\omega$}}} =
\overline{\omega}_s {\hat{\bf z}} \ . \eqno (3.24)$$
The vorticity magnitude is thus $\mid \overline{\omega}_s
\mid$.

Hence,  the twist of the world lines of static observers inside the shell
is simply equal to the dragging angular velocity $\overline{\omega}$, and
thus is increasing as the shell is collapsing. 

The vector of the angular velocity of rotation of the non-inertial
frame, ${\mbox{\boldmath $\omega$}}_s$, can also be calculated from
the gravitational vector potential (3.21): $\overline{\omega}^j =
\overline{\epsilon}^{jkl} \partial_k \overline{A}_{l}^{(g)}$.  In the
frame (3.19) only $\overline{A}_3^{(g)} = - \overline{\omega}_s r^2
{\rm sin}^2 \theta$ is non-vanishing; we find again
$\overline{\omega}^j$ given by (3.23).

In the above discussion we treated the regions outside and inside the
shell separately.  It is interesting to connect them by, in principle,
observable effects.  Lindblom and Brill \cite{3} considered how observers at
infinity will see a search light at rest in the inertial frame at the
center and what angular velocity of the matter of the shell they measure.  

Here we make a small addition to their considerations by calculating the
apparent motion of fixed stars observed by an inertial observer at the
center of the shell.

\subsection{The apparent motion of fixed stars}

The null geodesic motion of the star light emitted radially inwards in the
equatorial plane and received at the center is described by first order
differential equations.  These follow from angular momentum conservation
(zero in our case) and energy conservation.

Inside, the light goes on straight lines in the inertial frame:
$$\frac{d\overline{\varphi}_{\rm star}}{d\overline{t}} = 0 \; , \qquad \qquad
\frac{dr}{d\overline{t}} = -1 \qquad \qquad r \leq r_s \ . \eqno (3.25)$$
In terms of $\varphi$, the $\overline{\varphi}$ equation may be
written as in (2.7).  Outside,
$$\frac{d\varphi_{\rm star}}{dt} = \omega = \frac{2J}{r^3}\ ,  \qquad \qquad
\frac{dr}{dt} = - \left(1-\frac{2M}{r}\right) \ . \eqno (3.26)$$
The formal integration of these equations for $0 \leq r \leq
\infty$ gives a relation between $\varphi_{\rm star} (r=\infty) \equiv
\varphi_{\infty}$ and $\varphi_{\rm star} (r = 0,\overline{t})$ which we
shall
simply denote by $\varphi_{\rm star}$.

Thus,
$$\varphi_{\rm star} = \varphi_{\infty} +
\int^{\overline{t}}_{\overline{t}_{\rm ret}} \; \overline{\omega}_s
\; d\overline{t} \; + \; \int^{\infty}_{r_s(t_{\rm ret})} \;\;
\frac{\omega}{1-2 M/r} \; dr \ . \eqno (3.27)$$
Two photons arriving from the same place
($\Delta\varphi_{\infty} =
0)$ with a small time delay $\Delta t$ or $\Delta \overline{t}$ will be
seen in different directions; the change of direction is given by 
$$\Delta \varphi_{\rm star} = \overline{\omega}_s\;  \Delta \overline{t} -
\overline{\omega}_{s{\rm ret}} \; \Delta\overline{t}_{\rm ret} -
\left(\frac{\omega_s}{1- \frac{2M}{r_s}}\right)_{\rm ret} \Delta r_{s {\rm
ret}} \ . \eqno (3.28)$$
However, $\Delta \overline{t}, \; \Delta \overline{t}_{\rm ret}$
and
$\Delta r_{s{\rm ret}}$ are related by Eqs. (2.13) and (2.12).  One easily
finds that 
$$\Delta \overline{t}_{{\rm ret}} = \frac{1}{1 + \overline{V}_{\rm ret}} \Delta
\overline{t} \ , \eqno (3.29)$$
and
$$\Delta r_{s {\rm ret}} = \frac{\overline{V}_{\rm ret}}{1 +
\overline{V}_{\rm ret}} \Delta \overline{t} \ . \eqno (3.30)$$
If we insert those variations in (3.28) and use Eq. (2.11) for
$V$ and the fact that by definition -- see (2.7) and (2.8) -- 
$$\overline{\omega}_s \Delta \overline{t} = \omega_s \Delta t \ ,
\eqno (3.31)$$
we obtain for (3.28) an expression linear in $\Delta
\overline{t}$:
$$ \Delta \varphi_{\rm star} = \left[ \overline{\omega}_s (\overline{t}) - \left( \overline{\omega}_s {1 + V \over 1 + \overline{V}} \right)_{{\rm ret}} \right] \;
\Delta\overline{t} \ . \eqno (3.32)$$
Thus, from (3.32) it follows that
$$\overline{\omega}_{\rm star} = \frac{d\overline{\varphi}_{\rm
star}}{d\overline{t}} = - \left( \overline{\omega}_s  \;
\frac{1+V}{1+\overline{V}} \right)_{\rm ret} \ . \eqno (3.33)$$
This completes the derivation of Eq. (2.10) the properties of
which were
discussed in Section 2.

\section{The electromagnetic field of a collapsing rotating shell}

We shall be interested in the electromagnetic field of a charged
collapsing shell that may rotate rapidly.  This is a classical
problem, the solution of which is not entirely straightforward even in
flat space.  So we work it out.  We show here that we can construct
the vector potential for our problem in terms of the Green's function
for the spherical scalar wave equation.  That function is calculated
and used to derive the solution in the Appendix.

\subsection{Formal solution in terms of Green's function}

\subsubsection{Elements of the problem}

Let $A_0$ be the scalar potential and ${\bf A} = (A_k = -A^k)$ the
vector potential.  In the Lorentz gauge, Maxwell's equations for the
vector potential are given by
$$\Box {\bf A} = \nabla^2 {\bf A} - \partial^2_{\overline {t}} {\bf A} = -4 \pi
{\bf j} \ . \eqno (4.1)$$
The current ${\bf j}= (j^k)$ is that of a rotating shell of radius $r_s (t)$
and uniformly distributed charge $Q$:
$${\bf j} = {Q \over 4 \pi r^2_s} \left( \dot {r}_s {\hat {\bf r}} + 
{\overline {\mbox {\boldmath $\Omega$}}}_\tau \times {\bf r} \right) \delta (r-r_s) \ . \eqno (4.2)$$
${\overline {\mbox {\boldmath $\Omega$}}}_\tau = {\overline {\Omega}}_\tau {\hat {\bf z}}$ is the angular velocity of
the shell measured in proper time, ${\dot r}_s$ and ${\overline {\mbox {\boldmath $\Omega$}}} _\tau$
are related to the radial velocity $\overline {V} = d r_s/d \overline
{t}$ and the angular velocity of the shell $\overline {\Omega} = d
\overline {\varphi}/d \overline t$ as follows
$$\dot r _s = {\overline V \over \sqrt{1 - \overline V ^2 - r^2_s
\, \overline \Omega ^2 \sin ^2 \theta}}\ ,  \qquad {\overline {\Omega}}_\tau = {\overline
\Omega \over \sqrt{1 - \overline V ^2 - r^2_s \, \overline \Omega ^2 \sin
^2 \theta}}  \ \ . \eqno (4.3)$$

The scalar potential will be obtained by integration of the Lorentz
condition
$${\partial A_0 \over \partial \overline t} = {\mbox {\boldmath $\nabla$}} \cdot {\bf A} \ . \eqno (4.4)$$
The remaining gauge freedom, $A_0 \rightarrow A_0 + {\partial \zeta
\over \partial t}$, ${\bf A} \rightarrow {\bf A} + {\mbox {\boldmath $\nabla$}} \zeta$ with
$\Box \zeta = 0$, may be used to simplify the solution as we shall see.

The potentials are regular at $r = 0$ and vanish at infinity.  Initially
$(\overline t = 0)$ we take the radius of the shell at rest, $\dot r
_s (0) = 0$, and the angular velocity constant ${\overline {\Omega}} _0$.  Thus at
$$\overline t = 0 \, : \qquad A_0 = -{Q \over r_{s0}}   \quad r \leq r_{s0}\, ;
\qquad A_0 = - {Q \over r} \quad r \geq r_{s0}  \ , \eqno (4.5)$$
while
$$
\begin{array}{lll}{\bf A} = & {1 \over 3} Q  r_{s0}^{-1} \, {\overline{\mbox {\boldmath $\Omega$}}} _0 \times {\bf r}
\qquad &  r \leq r_{s0} \ ,  \\
&& \\
 {\bf A} = & {1 \over 3} Q r^2 _{s0}  \, r^{-3} \, 
{\overline{\mbox {\boldmath $\Omega$}}} _0 \times {\bf r} \qquad & r \geq r_{s0}\ .
\end{array} \eqno (4.6)
$$
So $${\bf B} = {\textstyle {2 \over 3}} Q r_{s0}^{-1}\, 
{\overline{\mbox {\boldmath $\Omega$}}} _0 \qquad r< r_{s0} \ . \eqno (4.7)$$
${\bf B} $ is dipolar with moment ${\textstyle {1 \over 3}} Q
r_{s0}^2 {\overline {\mbox {\boldmath $\Omega$}}} _0$.  

We have to solve (4.1) and (4.4) with initial conditions (4.5), (4.6)
and (4.7).

\subsection{Formal solution}

Suppose we have found the unique regular solution for the Green's
function $\chi (\overline t , r ; t', r')$ satisfying
$$\Box \chi = - \delta (r - r') \delta (\overline t - t') \ , \eqno (4.8)$$
and the conditions that $\chi$ tends to zero at $\overline t = \pm \infty
\, , \, r = \infty$.
Consider then the integral
$$\eta (\overline t, r, t') = \int ^{r_s (t')}_0 \chi dr' \ . \eqno (4.9)$$
Integrating (4.8) from $0$ to $r_s(t')$:
$$\Box \eta = - \Theta (r_s - r) \delta (\overline t - t') \ ,  \eqno
(4.10)$$
in which $\Theta (x)$ represents the step function $\Theta (x < 0)
= 0, \ \Theta (x>0) = 1$.  We deduce
$$\Box {\mbox {\boldmath $ \nabla$}} \eta = {\hat {\bf r}} \delta (r - r_s) \delta (\overline t
- t') \ . \eqno (4.11)$$
Multiplying both sides of (4.11) by ${Q \over r^{2}_s} \, {\dot r'}_s$
where the $\prime$ means that ${\dot r}'_s = {\dot r}_s (t')$ is a function of $t'$
-- and integrating over $t'$ between $\pm \infty$, we find
$$\Box {\mbox {\boldmath $\nabla$}} \zeta = {Q \over r^2_s} \dot r _s {\hat{\bf r}} \delta (r
- r_s) \ , \eqno (4.12)$$
where
$$\zeta = \int^{+ \infty}_{- \infty} {Q \over r'^2_s} \dot r '_s \eta
\, dt' \ . \eqno (4.13)$$
Multiplying both sides of (4.11) again by ${Q \over r'_s} {\overline{\mbox {\boldmath $\Omega$}}}
'_\tau \times$ and integrating again $t'$ between $\pm \infty$, we
find that
$$\Box \left( \int ^{+ \infty} _ {- \infty} {Q \over r'_s} \, {\overline{\mbox {\boldmath $\Omega$}}} '
_ \tau \times {\mbox {\boldmath $\nabla$}} \eta \,  dt' \right) = {Q \over r^2 _s}\, {\overline {\mbox {\boldmath $\Omega$}}} _\tau \times {\bf r} _s \delta (r - r_s) \ . \eqno (4.14)$$
If we add (4.12) to (4.14), we find that the right hand side of the
sum is minus the right hand side of Eq. (4.1) with (4.2).  The uniqueness of
$\chi$ guarantees that ${\bf A}$ is equal to minus the sum of the two
terms in (4.12) and (4.14) operated on by $\Box$:
$${\bf A} = - {\mbox {\boldmath $\nabla$}} \zeta - \int ^{+ \infty} _{- \infty} {Q \over
r'_s} {\overline{\mbox {\boldmath $\Omega$}}} ' _\tau \times {\mbox {\boldmath $\nabla$}} \eta \, dt' \ . \eqno (4.15)$$

We can now integrate (4.4) to obtain $A_0$.  Since $\eta$ is a
function of $r$, ${\mbox {\boldmath $\nabla$}} \eta = {\hat {\bf r}} {\partial \eta \over
\partial r}$ and, therefore, the divergence of the second term in
${\bf A}$ is identically zero.  Thus, (4.4) becomes
$${\partial A_0 \over \partial \overline t} = - \nabla^2 \zeta \
. \eqno (4.16)$$
However Eq. (4.10) reads
$$\nabla^2 \eta = {\partial ^2 \eta \over \partial \overline t ^2 } -
\delta (\overline t - t') \Theta (r_s - r) \ . \eqno (4.17)$$
With (4.17) and (4.13) we can readily evaluate the right hand side of
(4.16) and, integrating from $\overline t = 0$ to $\overline t$, we obtain
$$A_0 = A_0 (\overline t = 0, r) - \left . {\partial \zeta \over \partial
\overline t} + {\partial \zeta \over \partial \overline t}\right |
_{{\overline {t}}=0} + \int ^{\overline t} _ 0 {Q \over r^2_s} \dot r _s \Theta (r_s
- r) d \overline t \ . \eqno (4.18)$$ 

It should be remembered that $\dot r _s \neq dr_s/d \overline t =
\overline {V}$ and, therefore, the last integral is not readily
calculable unless $\overline V ^2 \ll 1$.  The electromagnetic
potential $A_\mu = (A_0, A_k)$ contains a term $\partial _\mu \tilde
\zeta$, with
$$\tilde \zeta = \zeta - \int ^{\overline t} d {\tilde t} \int ^{\tilde t} {Q \over
r^{\prime 2}_s} \dot r'_s \Theta (r'_s - r ) d t' \ . \eqno (4.19)$$
Since $\Box \tilde \zeta = 0$, as can easily be checked with the help
of (4.10), we may remove $\partial_\mu \tilde \zeta$ from the
solution.
It may be checked later, when we shall know the auxiliary function
$\chi$, that $\left . {\partial \zeta \over \partial \overline t }
\right | _{\overline t = 0}$ is zero.  We shall not show this
explicitly.  The solution we adopt is thus
$$A_0 (\overline t , r ) = A_0 (0, r) \ , \eqno (4.20)$$
and
$${\bf A} (\overline t , {\bf r}) = - \int ^{+ \infty}_{- \infty} {Q
\over r'_s} {\overline {\mbox {\boldmath $\Omega$}}} ' \times {\mbox {\boldmath $\nabla$}} \eta \, dt' \ . \eqno
(4.21)$$
We must now calculate $\chi$ to derive ${\mbox {\boldmath $\nabla$}} \eta = {\hat {\bf r}}
{\partial \eta \over \partial r}$.  This is done in the Appendix.

\section{Conclusions}

Our principal results are stated mathematically in Section 2.  Here we describe them qualitatively.

There are strong analogies between the gravitational effects of rotating and collapsing massive spherical shells and the electromagnetic effects of rotating and collapsing charged shells.  The precession of the inertial frame in the former is the analogue of the Larmor precession of a charged particle due to the magnetic field generated by the latter.  Thus the gravomagnetic field gives a good intuitive concept of the effects more normally ascribed to the ``dragging of inertial frames''.  However there are fundamental differences between the gravitational and electromagnetic problems because gravity carries no dipolar waves, so, in the natural gauge dipolar effects are instantaneous.  By contrast the electromagnetic field carries dipolar waves and when these are generated the fields suffer retardation.  Had we taken shells with quadrupolar distortions both the quadrupolar gravitational effects and the electromagnetic ones would have suffered retardation.

The formulae of Section 2 are expressed in Schwarzschild coordinates which make them simpler than those originally found in isotropic coordinates by Lindblom \& Brill \cite{3}.  A central observer at fixed orientation with respect to infinity may feel dizzy due to Coriolis force and will see the distant quasars rotating forwards with the whole sky, as the massive rotating sphere surrounding him falls inward.  (If it rotates without falling he merely sees the quasars in the `wrong' directions.)  Objects moving near him will be subject to both Coriolis and Euler forces and, as his orientation is not inertial, he needs a torque to keep him oriented.  By contrast a lazy inertial observer near the centre will rotate at a time dependent rate with respect to infinity, will suffer no Coriolis giddiness but will see the sky rotating backwards while the massive shell around him rotates forwards.  We distinguish between the fixed direction to a distant quasar in the world map at one time and its apparent direction from which the light is seen.  The latter is deviated by the gravomagnetic field (and in a time dependent way if the rotating sphere also falls).  

\section*{Appendix}

\subsection*{(a) Green's function and the vector potential}
Consider first, e.g., (4.8) for $\chi (t, r )$ which may be rewritten
$${\partial ^2 (r \chi) \over \partial \overline t ^2} - {\partial ^2
(r \chi) \over \partial r ^2} = r \delta (r - r') \delta (\overline t
- t') \ . \eqno ({\rm A.1})$$
A solution {\it regular} at the origin for $0 < r < r'$ is of the
general d'Alembert form with ingoing and outgoing waves:
$$r \chi = F_1 (\overline t - r ) - F_1 (\overline t + r) \qquad r <
r' \ . \eqno ({\rm A.2})$$
For $r>r'$ we are interested in outgoing waves only; $r \chi$ has then
to be of the form
$$r \chi = F_2 (\overline t - r) \qquad r>r'  \ . \eqno ({\rm A.3})$$
$\chi$ must be continuous at $r = r'$, so
$$F_2 (\overline t - r') = F_1 (\overline t - r') - F_1 (\overline t +
r') \ . \eqno ({\rm A.4})$$
Replacing $\overline t$ by $\overline t - r + r'$ gives
$$F_2 (\overline t - r) = F_1 (\overline t - r) - F_1 (\overline t - r
+ 2 r') \qquad r>r' \ . \eqno ({\rm A.5})$$
To obtain $F_1$, we integrate (A.1) across $r = r'$.  The time
derivatives disappear in the process and we find that
$${\lim \atop \varepsilon \rightarrow 0} \ \ \left[ - \left. {\partial (r
\chi) \over \partial r} \right | ^{r \rightarrow r' + \varepsilon }_{r
\rightarrow r' - \varepsilon}  \right] = \ r' \delta (\overline
t - t') \ . \eqno ({\rm A.6})$$
With the left hand side calculated with (A.2) inside $(r < r')$ and
(A.3),(A.4) outside $(r>r')$, we find that the derivative $F_1'$ of
$F_1$ with respect to its argument satisfies the relation
$$-2 F'_1 (\overline t +r') = r' \delta (\overline t - t') \ . \eqno
({\rm A.7})$$ Integrating $F'_1$ over $\overline t$ between $- \infty$ and
$\overline t$, assuming $F_1 (- \infty) = 0$, we obtain
$$F_1 (\overline t + r') = - {r' \over 2} \Theta (\tilde t)\ ,  \qquad
\tilde t \equiv \overline t - t' \ . \eqno ({\rm A.8})$$

We shall now insert $F_1$ with the correct arguments into (A.2) and (A.4)
to obtain $\chi (\overline t , r ; t', r')$:
$$\chi = {r' \over 2r} \left [ \Theta (\tilde t + r - r') - \Theta (\tilde t -
r - r') \right ] \qquad r<r' \ , \eqno ({\rm A.9})$$
and
$$\chi = {r' \over 2r} \left[ \Theta (\tilde t - r + r') - \Theta (\tilde
t - r - r') \right ] \qquad r>r' \ . \eqno ({\rm A.10})$$
For $\tilde t < 0$, $\chi = 0$ but for $\tilde t>0$, $\chi$ has the
simple form illustrated in Figure 1.

\begin{figure}
\centerline{\psfig{figure=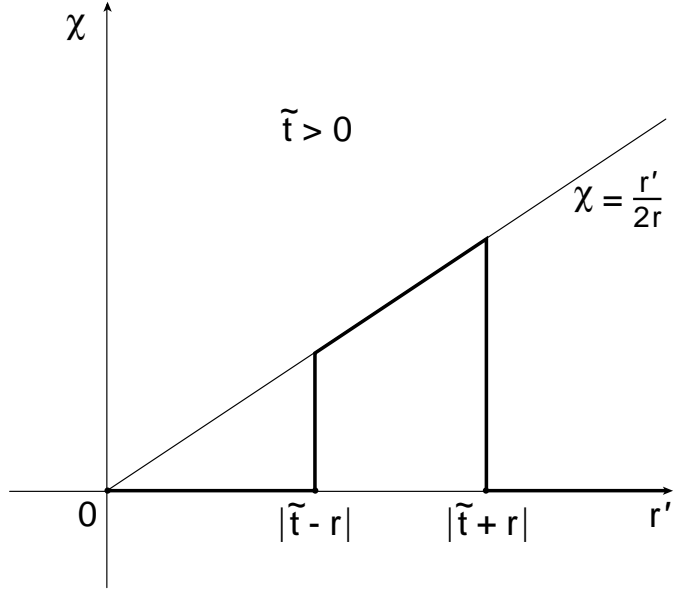,width=3.5in}}
\caption{The function $\chi (r')$ for $\tilde t > 0$.}
\end{figure}

With (A.9), (A.10), or easier from Figure 1, we can calculate $\eta
(\tilde t, r, t')$ defined in Eq. (4.9).  First notice that for
$\tilde t<0 \ , \ \eta = 0$ and for $\tilde t > 0 \ , \break \ \eta (r'_s
< | \tilde t - r | ) = 0$.  The function is not equal to zero and
assumes different values in two different intervals of values for $r'_s$: for 
$$| \tilde t - r | < r'_s < | \tilde t + r | \, , \quad \tilde t > 0
 \, , \qquad \quad \eta = {1 \over 4r} \ [r^{\prime 2}_s - (\tilde t - r)
 ^2] \ , \eqno ({\rm A.11})$$ and for
$$r'_s > | \tilde t + r |\, , \quad \tilde t > 0\, , \qquad \quad \eta = \tilde t \
. \eqno ({\rm A.12})$$

We thus find that ${\partial \eta \over \partial r}$ is only
different from zero if $|\tilde t - r | \leq r'_s \leq | \tilde t + r
| $ and $\tilde t > 0 $:
$$\tilde t > 0 \ , \ | \tilde t - r | < r'_s < \tilde t  + r
\qquad {\partial \eta \over \partial r } = { 1 \over 4r^2} \, (\tilde t ^2
- r^2 - r^{\prime 2}_s ) \ . \eqno ({\rm A.13})$$
Remembering that $\tilde t = \overline t - t'$, we deduce from ({\rm A.13})
the intervals of $t'$ within which ${\partial \eta \over \partial r}
\neq 0$:
$${\partial \eta \over \partial r} \neq 0 \left \{ {{\rm inside} \ \ \
\ r < r_s (t') \ \ {\rm for} \ \ \overline t - (r'_s + r) \leq t' \leq
\overline t - (r'_s - r) \atop {\rm outside} \ \ r > r_s (t') \ \ {\rm
for} \ \ \overline t - (r'_s + r) \leq t' \leq \overline t - (r- r'_s)}
\right. \ .\eqno ({\rm A.14})$$

With (A.14) and ${\partial \eta \over \partial r}$ given in (A.13), we
can calculate ${\bf A}$ given in (4.21).  The lower bound of $t'$
inside and outside of the shell is the same; it is a retarded time
which we denote by $\overline t^-$,
$$\overline t ^- = \overline t - r - r_s (\overline t ^-) \qquad r <
r_s (\overline t ^-) \ . \eqno ({\rm A.15})$$
There are, however, two different upper bounds $\overline t_{{\rm up}}$ which
are both advanced times.  Inside the shell,
$$\overline t_{{\rm up}} = \overline t^+ = \overline t + r - r _s
(\overline t ^+)\ , \qquad r<r_s (\overline t ^+) \ . \eqno ({\rm A.16})$$
Outside,
$$\overline t _{{\rm up}} = \overline t ^* = \overline t - r + r _s
(\overline t ^*)\ , \qquad r>r_s (\overline t ^*) \ . \eqno ({\rm A.17})$$
The vector potential ${\bf A}$ as given in (4.21) contains $\mbox {\boldmath $\nabla$}
\eta = {\partial \eta \over \partial r} {\hat {\bf r}}$.  We replace
${\partial \eta \over \partial r}$ by its expression (A.13) and obtain
$${\bf A} = - {Q \over 4r^3} \int ^{\overline t_{{\rm up}}} _ {{\overline t}^-} {1 \over
r'_s} {\overline {\mbox {\boldmath $ \Omega$}}} ' _ \tau \times {\bf r} \; [(\overline t - t')^2 - r^2 -
r^{\prime 2}_s ]\; dt'  \ . \eqno ({\rm A.18})$$

\subsection*{(b) The electromagnetic fields near the center}

The scalar potential within the shell is a constant $A_0 = -Q/r_{s0}$.
As a result, both ${\bf E}$ and ${\bf B}$ are determined by ${\bf A}$.
We shall here expand the expression of ${\bf A}$ near the center of
the shell in powers of $x^k$.  In this way we shall clearly see that
${\bf E}, {\bf B}$ are not uniform, their dependence on $\overline t$
is entirely through the motions of the shell which we may specify how
we like.  To make the expansion we first rewrite (A.18) in the form
$${\bf A} = {Q \over 4 r^3} {\mbox {\boldmath $\psi$}} \times {\bf r}
\ . \eqno ({\rm A.19})$$ Thus, inside,
$${\overline {\mbox {\boldmath $\psi$ }}} = - \int^{t^+} _{t^-}
{{\overline {\mbox {\boldmath $\Omega$}}} '_\tau \over r'_s} \; [(\overline t -
t')^2 - r^2 - r^{\prime 2}_s ] \; dt' \ . \eqno ({\rm A.20})$$
Eqs. (A.19) and (A.20) are the equations quoted in (2.26) and (2.27).

To expand $\mbox {\boldmath $\psi$}$ in powers of ${\bf r}$ we shall
have to calculate the derivatives of $t^+$ and $t^-$ with respect to
$t$.  These are readily obtained from (A.15) and (A.16) in terms of
the radial velocity of the shell
$$\overline V = {dr_s \over d \overline t} \ , \eqno ({\rm A.21})$$
calculated at $\overline t ^+$ and $\overline t^-$; thus,
$${d \overline t ^+ \over d r} = {1 \over 1 + \overline V ^+} \ \ , \
\ {d \overline t ^- \over dr} = - {1 \over 1 + \overline V ^-} \ , \eqno
({\rm A.22})$$ where $\overline V ^+ = \overline V (\overline t^+)$ and
$\overline V ^- = \overline V (\overline t ^-)$.  The first derivative
of $\mbox {\boldmath $\psi$}$ with respect to $r$ is obtained with the
help of (A.22) and is of the form
$${d {\mbox {\boldmath $\psi$}} \over dr} = 2 r {\mbox {\boldmath
$\nu$}} \ , \eqno ({\rm A.23})$$ with
$${\mbox {\boldmath $\nu$}} = {{\overline {\mbox {\boldmath $\Omega$}}} ^+_\tau \over 1 + \overline V ^+} - {{\overline{\mbox {\boldmath
 $\Omega$}}} ^- _ \tau \over 1 + \overline V ^-} + \int ^{\overline t ^+}
_{\overline t ^-} {{\overline{\mbox {\boldmath $\Omega$}}} '_\tau \over r'_s} dt' \ . \eqno
({\rm A.24})$$
The method for deriving successive derivatives is obtained by
calculating $\nabla ^2 (r^2 {\bf A} )$.  One easily can find the
following identity from (A.20) and (A.21)
$$- {2 \over Q} \left [ 6 {\bf A} + 4 ({\bf r} \cdot {\mbox {\boldmath
$\nabla$}} ) {\bf A} + r^2 \, \nabla ^2 \, {\bf A} \right] = {{\bf r}
\over r} \times \left(3 {\mbox {\boldmath $\nu$}} - {{\mbox {\boldmath
$\psi$}} \over r^2} + r {d {\mbox {\boldmath $\nu$}} \over dr} \right)
\ . \eqno ({\rm A.25})$$ To expand ${\bf A}$ in powers of ${\bf r}$, we need
the patience to calculate higher order derivatives of (A.25) and
evaluate each of them at ${\bf r} = 0$.  The result to order 3 is as
follows:
$${\bf A} = {Q \over 6} \ \left [ \left ( {d {\mbox {\boldmath $\nu$}} \over d r}
\right)_{r = 0} + \left ( {d^3 {\mbox {\boldmath $\nu$}} \over d r^3} \right)_{r = 0}
{r^2 \over 10} \right ] \times {\bf r} + {\bf 0} (r^5) \ . \eqno
({\rm A.26})$$
The ${\bf E}$ and ${\bf B}$ fields are both $t$ and ${\bf r}$ dependent.
To lowest order, ${\bf E}$ and ${\bf B}$ near the center are given by
$${\bf B} \simeq ({\mbox {\boldmath $\nabla$}} \times {\bf A} ) _{r=0} = {Q \over 3}
\left ( {d {\mbox {\boldmath $\nu$}} \over dr} \right)_{r=0} \, = {\bf B}_c \ , \eqno ({\rm A.27})$$
and
$${\bf E} \simeq - {1 \over 2} \; {d {\bf B} \over dt} \times {\bf r} \
. \eqno ({\rm A.28})$$
The magnetic field ${\bf B}_c = B_c {\hat {\bf z}}$ can be calculated from (A.24),
$$B_c = {2 Q \over 3} \; \left[ {1 \over 1 + \overline V}\;  {d \over
d \overline t} \left ( {{\overline \Omega}_\tau \over 1 + \overline V} \right) + {{\overline \Omega} _\tau/r_s
\over (1 + \overline V)^2} \right]_{\overline t _{{\rm ret}}} \ , \eqno ({\rm A.29})$$
${\overline t}_{{\rm ret}}$ is the time at which the shell had a radius $r_s
(\overline t _{{\rm ret}})$:
$$\overline t - r_{s{\rm ret}} = \overline t_{{\rm ret}} \ . \eqno
({\rm A.30})$$
For a shell of constant radius $r_{s0}$ and rotating at constant angular
velocity $\Omega_0$, $B_c = B_0$ given in (4.7).  

Equation (A.29) with ${\overline {\Omega}}_\tau$ expressed in terms of ${\overline {\Omega}}$, see (4.3), is equation (2.33) when ${\overline {\Omega}}^2$ terms are neglected.

\end{document}